\documentclass[balance,upint,subscriptcorrection,varvw,mathalfa=cal=boondoxo,pdf-a,colorlinks]{asmeconf}

\hypersetup{%
	pdfauthor={Giuseppe Bruni},									  
	pdftitle={C(NN)FD - Deep Learning Modelling of Multi-Stage Axial Compressors Aerodynamics},                  
	pdfkeywords={},
	pdfsubject = {},			  
	pdflicenseurl={}
}

\begin{document}


\ConfName{Pre-print submitted to the Proceedings of the ASME Turbo Expo 2024 \linebreak Turbomachinery Technical Conference and Exposition}
\ConfAcronym{GT2025}
\ConfDate{June 17 - 19, 2025} 
\ConfCity{Memphis, USA} 
\PaperNo{-}


\title{C(NN)FD - Deep Learning Modelling of Multi-Stage Axial Compressors Aerodynamics} 

\SetAuthors{%
	Giuseppe Bruni$^{1,2}$\CorrespondingAuthor{giuseppe.bruni@siemens-energy.com}, 
   Sepehr Maleki$^{2}$
	Senthil K. Krishnababu$^{1,2}$
	}

\SetAffiliation{1}{Siemens Energy Industrial Turbomachinery Ltd, Lincoln, United Kingdom}
\SetAffiliation{2}{Lincoln AI Lab, University of Lincoln, United Kingdom}

\maketitle





\begin{abstract}
   The field of scientific machine learning and its applications to numerical analyses such as Computational Fluid Dynamics (CFD) has recently experienced a surge in interest. While its viability has been demonstrated in different domains across a wide range of engineering applications, it has not yet reached a level of robustness and scalability to make it practical for industrial applications in the turbomachinery field. The highly complex, turbulent, and three-dimensional flows of multi-stage axial compressors for gas turbine applications represent a remarkably challenging case. This is due to the high-dimensionality of the regression of the flow-field from geometrical and operational variables, and the high computational cost associated with the large scale of the CFD domains. This paper demonstrates the development and application of a generalized deep learning framework for predictions of the flow field and aerodynamic performance of multi-stage axial compressors, which is also potentially applicable to any type of turbomachinery. A physics-based dimensionality reduction unlocks the potential for flow-field predictions for large-scale domains, re-formulating the regression problem from an unstructured to a structured one and reducing the number of degrees of freedom. The relevant physical equations are used to define a multi-dimensional physical loss function, which considers the flow field, gradient of the flow field, radial distributions and integral quantities. Compared to “black-box” approaches, the proposed framework has the advantage of physically explainable predictions of overall performance, as the corresponding aerodynamic drivers can be identified on a 0D/1D/2D/3D level. An iterative architecture is employed, improving the accuracy of the predictions, as well as estimating the associated uncertainty at each level of the predictions. A quantifiable measure of the confidence in resolving the relevant flow features is provided, without a significant increase in computational cost.  The model is trained on a series of dataset including manufacturing and build variations, as well as different geometries, compressor designs and operating conditions, representative of typical aerodynamic assessments. This demonstrates the capability to predict the flow-field and the overall performance in a generalizable manner, for different geometries and compressors across their maps, with accuracy comparable to the CFD benchmark.
\end{abstract}

\begin{nomenclature}
   \entry{CNN}{Convolutional Neural Network}
   \entry{PCA}{Principal Component Analysis}
   \entry{VGV}{Variable Guide Vane}
   \entry{$\dot{m}$}{Corrected mass-flow}
   \entry{$\eta_p$}{Polytropic Efficiency}
   \entry{$PR$}{Pressure Ratio}
   \entry{$P_t$}{Total Pressure}
   \entry{$T_t$}{Total Temperature}
   \entry{$\rho$}{Density}
   \entry{$V_x$}{Axial Velocity}
   \entry{$V_t$}{Tangential Velocity}
   \entry{$V_r$}{Radial Velocity}
   \entry{$R^2$}{Coefficient of Determination}
   \entry{$MAE$}{Mean Absolute Error}
   \entry{$MSE$}{Mean Squared Error}
   \entry{$\mathcal{L}$}{Loss Function}
\end{nomenclature}

\section{Introduction}

Advancements in modeling and numerical methods have progressively improved accuracy and reduced computational costs of CFD analyses, making them an integral part of industrial design processes for turbomachinery components in gas turbine applications. Regardless, the application of CFD in the design and analysis of axial compressors and turbines remains computationally expensive and time-consuming. This is especially relevant in the case of multi-stage axial compressors, due to the large computational domain associated with the high number of stages. As a result, there is a growing interest in developing alternative methods that can provide accurate predictions of the flow field and aerodynamic performance, with reduced computational cost. These methods can then be deployed to support the design and analysis activities for a variety of aerodynamic assessments, including predicting the effect of variations in: tip clearance, surface roughness, in-tolerance geometries, Variable Guide Vanes (VGVs) settings, operating conditions, or a combination of these for a given compressor designs. Moreover, in case of design activities, surrogate models can be used to provide quick assessment for manual design iterations, or supporting optimisation activities by performing low-fidelity analyses in parallel to high-fidelity ones. Conventional surrogate models, map relationships between specific input and output variables, and have to be built from scratch for each optimisation setup. While surrogate models have historically been developed using different approaches with varying degrees of complexity, developments in machine learning methods have unlocked the potential for more detailed, accurate and re-usable models. The application of scientific machine learning to numerical analyses such as Computational Fluid Dynamics (CFD) has recently experienced a surge in interest and its viability has been demonstrated in different domains across a wide range of engineering applications, both in academia as well as in industry. However, it has not yet reached a level of robustness and scalability to make it practical for industrial applications in the turbomachinery field. The highly complex, turbulent, and three-dimensional flows of multi-stage axial compressors for gas turbine applications represent a remarkably challenging case for the application of machine learning models for aerodynamic predictions. This is due to the high-dimensionality of the regression of the flow-field from geometrical and operational variables, and the high computational cost associated with the large scale of the CFD domains, which limits the amount of training data available compared to typical machine learning applications.

\section{Literature Review}
In recent years, there has been a significant amount of research into the utilisation of machine learning techniques for turbomachinery applications. These include predicting the aerodynamic performance due to tip clearance variations \cite{Krishnababu2021} and design modifications \cite{Pongetti2021}, as well as aeromechanic predictions for both forced response \cite{Bruni2022}, flutter \cite{He2022}, and for multifidelity aeromechanical design frameworks \cite{Remy2024}. Output variables are often predicted directly from the input variables with a "black box" approach. Depending on the complexity of the application, these methodologies have varying degrees of accuracy and generalisability. However, the ability to accurately predict variations in target variables for turbomachinery in a manner that is both reliable and generalisable, is contingent upon the accuracy of flow field predictions. Precise flow field predictions and the application of relevant physical equations that govern the dynamics are required to calculate the overall performance figures of interest \cite{Bruni2023} \cite{Bruni_Maleki_Krishnababu_2025}. The potential benefits of using machine learning approaches for predicting full flow fields, have been demonstrated in some simplified cases in the literature for 2D aerofoils on both cartesian \cite{Thuerey2020}, structured  \cite{chen2023} and unstructured \cite{Kashefi2021} grids. Convolutional Neural Networks (CNN) have been demonstrated to predict the flow field downstream of a single fan row, which was then used for noise predictions applications \cite{Li2023}, and for overall performance predictions \cite{Fesquet2024}. A similar approach was used also by Rao et. al. \cite{Rao2023} for the semantic segmentation of an aero-engine intake using U-Net and U-Net++ architectures. Early attempts to apply structured approaches through voxelization of a fan stage 3D domain were found to have limited accuracy, especially for estimating overall performance figures \cite{Aulich2019}. The limitations of structured approaches based on convolutional neural networks for turbomachinery have led to a growing interest in graph neural networks \cite{Harsch2021} \cite{Pfaff2021}, which have been applied to 3D turbomachinery, on a single row compressor \cite{Perrone2022} and turbine \cite{Li2022}, as well as for unsteady predictions \cite{Stronisch2024}. Even when adopting a multi-scale approach to address memory limitations \cite{Stronisch2024}, challenges related to scalability are self-evident, and the limited flexibility due to the requirement of using CFD grids limit their practicality for industrial application. In a recent work, a transformer architecture was used to approximate the 3D flow field of multi-stage axial compressors for optimisation surrogate model applications \cite{Aulich2024}. The use of different sub-models for each blade-row and randomly selecting nodes from the CFD mesh for interpolation can potentially address some of the limitations previously discussed, and the approach is demonstrated to provide a useful low-fidelity surrogate model for optimisations. However, the predictions of overall performance metrics such as efficiency were found to be inaccurate, as minor changes in total temperature and pressure at the compressor inlet and outlet have a significant effect on  efficiency, but a negligible effect on the loss function of the model, due to the pure data-driven supervised learning approach used. Physics-informed neural network have recently gained popularity \cite{Raissi2019} and while presenting challenges with implementations for high-dimensional problems, are a potential solution to improve the physical consistency of machine learning based predictions.

\section{Novel contributions} \label{sec:novelty}

\begin{figure*}[t]
   \centering
      \includegraphics[width=\textwidth]{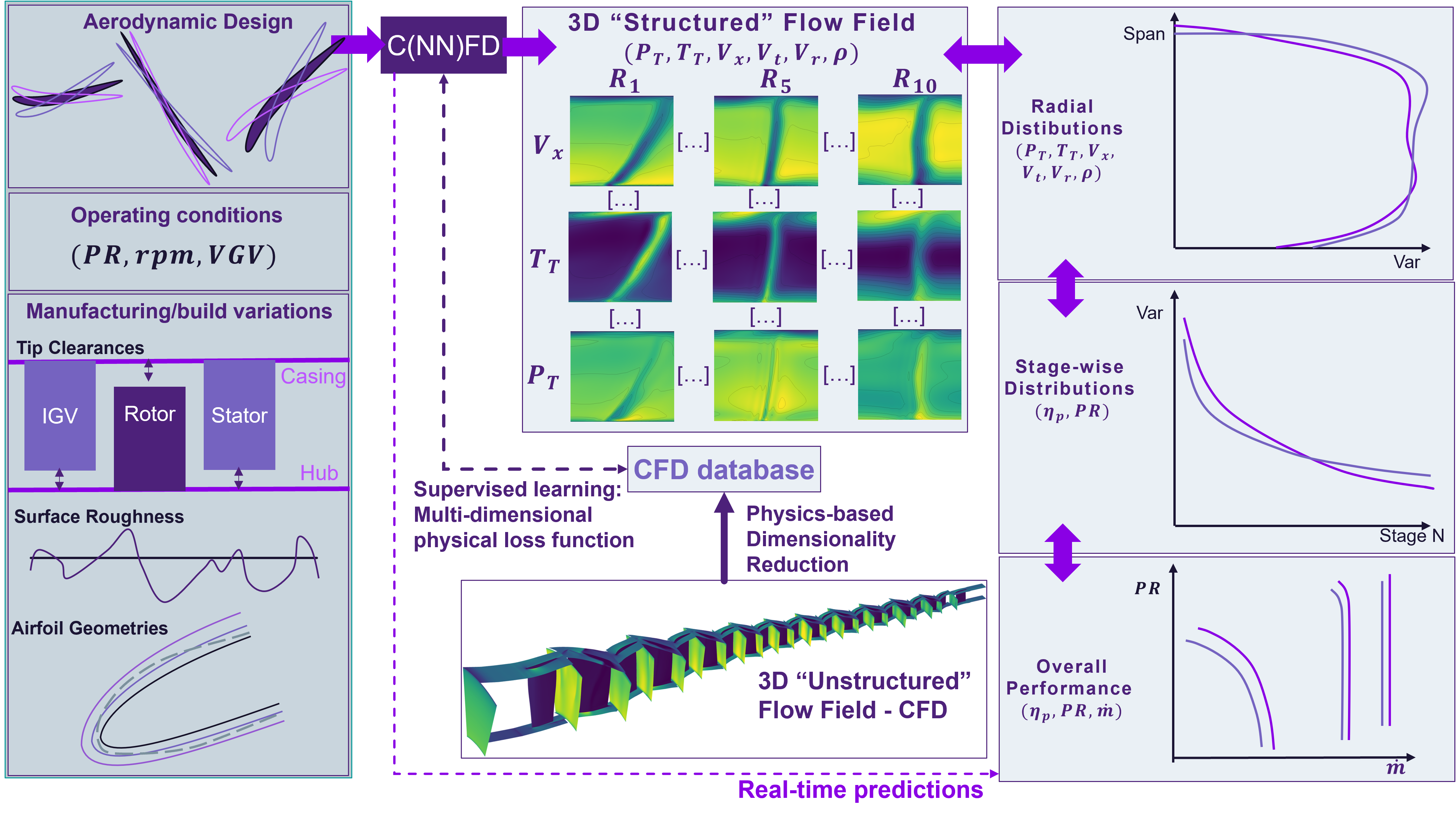}
      \caption{Overview of the C(NN)FD framework}
      \label{fig:graphical_abstract}
\end{figure*}

The challenges associated with using machine learning for compressor aerodynamic predictions are the size of the computational domain associated with typical CFD grids, the limited amount of data typically available for training, the complexity of the flow-field to be predicted, and the significant effect of small errors at the inter-row locations on the overall performance predictions. To address these challenges, the authors have introduced in previous work \textit{C(NN)FD}, a deep learning framework for predictions of flow fields and aerodynamic performance of turbomachinery components \cite{Bruni2023} \cite{Bruni_Maleki_Krishnababu_2025}. In addition to the previously proposed physics-based dimensionality reduction, the current work introduces to the methodology a multi-dimensional physical loss function, an iterative architecture and uncertainty estimations. Through the use of transfer learning, the application is then extended to a wider range of use cases, including manufacturing and build variations, the modelling of different operating conditions and geometrical design variations. While the focus of the current work is on multi-stage axial compressors, the framework is directly applicable to any other turbomachinery application, such as axial turbines. An overview of the C(NN)FD framework is presented in \autoref{fig:graphical_abstract}.

\subsection{Physics-based dimensionality reduction}
The C(NN)FD framework \cite{Bruni2023} \cite{Bruni_Maleki_Krishnababu_2025} employs a pre-processing step to extract only the relevant engineering data from the 3D CFD computational domain, such as inter-row axial contours, radial distributions, stage-wise performance, and overall performance. These provide sufficient information to the aerodynamicist for the majority of design and analyses activities. Once the relevant engineering data has been extracted from the computational domain, it can be interpolated onto a simplified grid with a resolution that is deemed acceptable for the application in question. The regression problem is therefore reformulated, from predicting the relevant variables on an unstructured computational grid, to a structured grid two orders of magnitude smaller. Only selected fundamental variables are predicted, while the derived ones are computed using the appropriate physical equations, significantly reducing the computational cost without sacrificing accuracy. This physics-based dimensionality reduction, unlocks the potential to use supervised learning approaches with convolutional neural networks for turbomachinery applications, as it re-formulates the regression problem from an un-structured to a structured one, as well as reducing the number of degrees of freedom. 

\begin{figure*}[t]
   \centering
      \includegraphics[width=\textwidth]{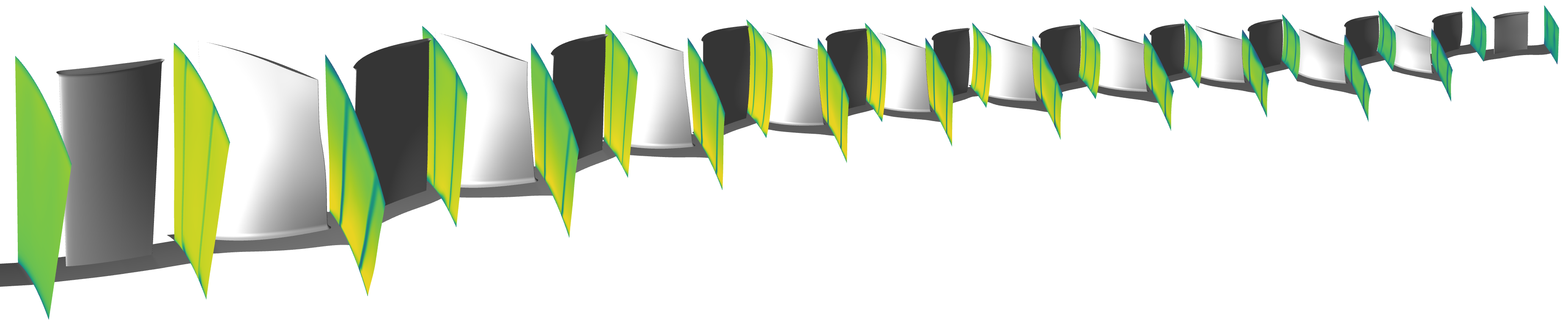}
      \caption{Overview of the CFD domain and axial velocity contours at the mixing-plane locations}
      \label{fig:domain_paper}
\end{figure*}

\subsection{Multi-dimensional physical loss function}
The C(NN)FD framework provides physical explainability to the predictions, as the corresponding aerodynamic drivers are identified by providing predictions for the 3D "structured" flow-field, as well as 2D radial distributions at all the inter-row locations, 1D row-wise distributions, stage-wise distributions, and 0D overall performance \cite{Bruni_Maleki_Krishnababu_2025}. The accuracy of the 0D/1D/2D/3D predictions is improved by using a physics-informed supervised learning approach \cite{Raissi2019}, defining a multi-dimensional physical loss function, which considers the flow field, spatial gradient of the flow field \cite{Fesquet2024}, radial distributions and integral quantities. The novelty of the proposed multi-dimensional physical loss function lies in its ability to incorporate additional physical constraints into the training process of 3D turbomachinery applications. By calculating the radial distributions and integral quantities, and including them in the loss calculation, the model is encouraged to produce predictions that are not only accurate at the flow field level but also consistent with the underlying physical principles. For instance, mass conservation is not imposed explicitly through the corresponding governing equation in the loss function, but rather implicitly by constraining the integral quantities at each inter-row locations to match the CFD solution. By incorporating physical constraints, the model is also better able to generalize to unseen data and different operating conditions, requiring smaller datasets for supervised training, which in turn enables addressing typical aerodynamic tasks in more complex and industrially relevant applications. While purely data-driven approaches typical of supervised learning would require large amount of data to learn the solution manifold, adopting appropriate physics-informed approaches unlocks the potential to scale to more complex and industrially relevant applications, addressing various typical aerodynamic assessment activities.  

\subsection{Iterative architecture and uncertainty estimation}
The C(NN)FD framework employs an iterative architecture, in which the flow field of a baseline configuration of reference is provided as input to the network. The output of the network is then fed through the input recursively across an inner loop. This training strategy is demonstrated to improve the accuracy of the predictions \cite{Mosinska2018}, as well as providing an estimate of both the epistemic and aleatoric uncertainties of the predictions \cite{Durasov2024}.

\subsection{Transfer Learning}
The C(NN)FD framework adopts a meta-learning approach by training a model on a variety of learning tasks, such that it can solve new learning tasks using only a small number of training samples \cite{Chelsea2017}. The model is trained on a wide dataset available from the industrial partner, including CFD data from various aerodynamic assessments of different compressor designs, operating conditions, variable guide vanes schedules, manufacturing and build variations, and geometries. This provides a pre-trained base model, which is then fine-tuned for the specific application of interest required for a given application, thus significantly reducing the amount of data needed to address new applications that were not covered by previous activities.

\section{Methodology}

The focus of the work is on multi-stage industrial axial compressors, and various configurations representative of different compressor designs are considered. For each configuration, the effect of  variations in tip clearance, surface roughness, geometry within the tolerance range (i.e. in-tolerance geometries), Variable Guide Vanes (VGVs) settings, operating conditions and airfoil designs are considered, depending on the task.

\begin{figure*}[t]
   \centering
      \includegraphics[width=0.99\textwidth]{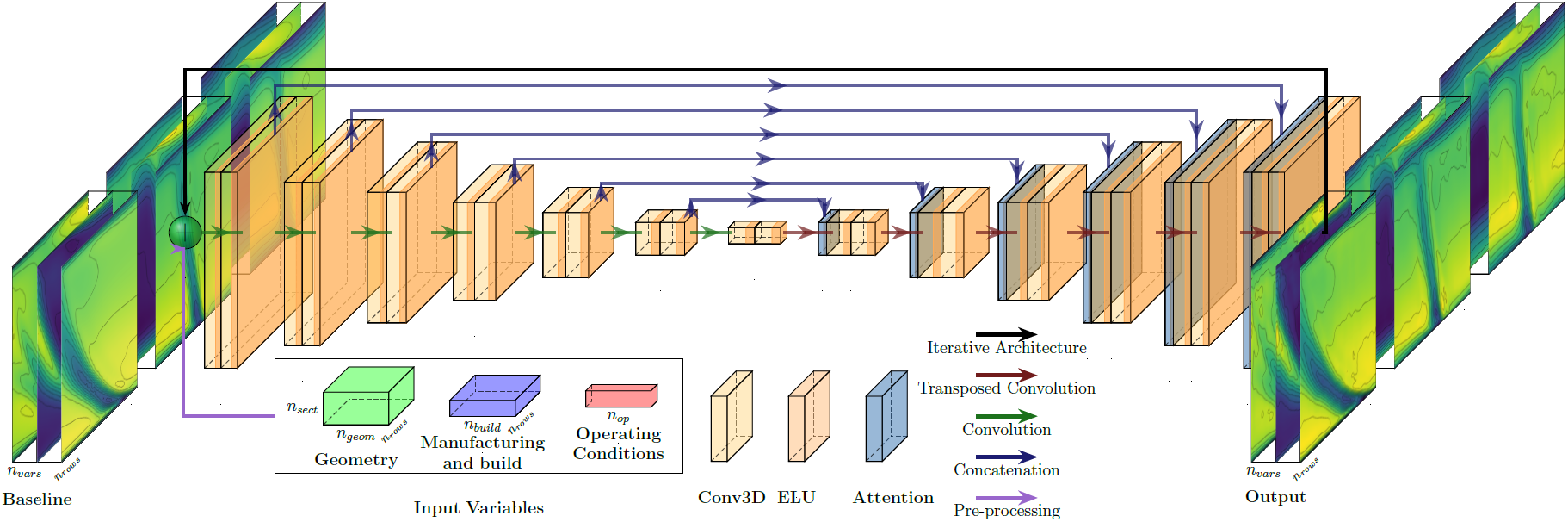}
      \caption{C(NN)FD architecture overview}
      \label{fig:architecture}
\end{figure*}

\subsection{Data Generation}
The ground-truth data used for training are the CFD results for the configuration of interest.  An overview of the computational domain is shown in \autoref{fig:domain_paper}, showing both the geometry of the airfoils as well as the axial velocity contours at the mixing plane locations. The computational mesh was generated using Numeca AUTOGRID5, and the CFD solver used was Trace, with SST $K-\omega$ turbulence model. More details on the computational setup have been provided in previous publications \cite{Bruni2022}, as well as its validation against engine test data \cite{Bruni2024CFD}. The quantification of the accuracy of the CFD simulations in comparison to test data is outside of the scope of the current work. The aim of the proposed architecture is to provide predictions of the flow field and overall performance of the compressor with accuracy similar to that of the CFD model, but without the associated computational cost. Each CFD calculation was executed using automated and parallelized processes, taking less than 90 minutes on 72 CPUs. The datasets used for this study includes variations in tip clearance, surface roughness, in-tolerance geometries, Variable Guide Vanes (VGVs) settings, operating conditions and airfoil designs for each configuration. This extends the dataset that was considered in previous work, which was focused on manufacturing and build variations for a given compressor design and operating point \cite{Bruni_Maleki_Krishnababu_2025}. Even if the computational cost required to generate these datasets for supervised training is potentially high, it should be noted that all the data was already available from internal work of the industrial partner. It is envisioned that the majority of the data used for training of this framework will be based on available data, with additional carefully curated datasets to be generated only when required, for wider generalisability. The use of a supervised learning approach is appropriate for industrial applications, as typically gas turbine manufacturers have access to a significant historical database of CFD calculations from previous analytical and design activities. In case of new applications for which historical data is not available, it's advisable to favour physics-informed models, as the computational cost associated with generating the data from scratch would render supervised models impractical.

\subsection{Data Pre-Processing}
The CFD results are pre-processed to reduce the dimensionality of the problem, without any loss of relevant data. First, only six variables $n_{vars}$ are extracted, namely, Total Pressure ($P_t$), Total Temperature ($T_t$), Axial Velocity ($V_x$), Tangential Velocity ($V_t$), Radial Velocity ($V_r$), and Density ($\rho$). The flow condition at each mesh node is fully described by the six variables and can be used to calculate all other relevant flow variables in the CFD solution. Second, only the inter-row axial locations shown in \autoref{fig:domain_paper} are considered. Third, as the methodology aims to be generalisable, the data is interpolated for a given passage at a fixed circumferential location, irrespective of the periodic boundary definitions, number of points and discretization used in the mesh. As the location of the periodic boundaries and distribution of the nodes can vary depending on the meshing approach or geometry, this approach is not limited to a given mesh or geometry, as typical in most surrogate models developed for optimisations.  While the CFD mesh for each simulation comprises generally of more than $10^7$ nodes, the new pre-processed domain contains less than $10^5$ nodes, while including all the relevant information. The new 3D flow field now consists of only $n_{rows}$ axial planes with a $n_{rad} \times n_{tang}$ grid. The data is stored as a tensor with a channel-first convention and a shape of $(n_{vars} \times n_{rows} \times n_{rad} \times n_{tang})$, where the dimensions represent respectively the number of variables, number of axial locations, radial nodes in the mesh, and tangential nodes in the mesh. The axial contours are generally only used for detailed analyses and the radial profiles of the variables of interest are obtained by either mass-flow or area averaging \cite{CumpstyAveraging} in the circumferential direction. These radial profiles of each variable are then averaged radially to obtain 1D integral quantities, which are used to calculate stage-wise performance of the variables of interest, such as pressure-ratio ($PR$), and polytropic efficiency ($\eta_p$). In addition to the aerodynamic performance of the stage, it is possible to calculate the overall performance of the compressor. Predicting the axial contours, radial distributions and stage-wise performance in addition to the overall performance is critical for implementation of the model within design and analysis processes, as it provides physical explainability to the effect of the input variables on the overall performance, by identifying the corresponding flow features.

\subsection{The C(NN)FD framework}

The tip clearance and surface roughness values, as well as the relevant geometry design parameters and operating conditions of interest are fed as input to \textit{C(NN)FD}, together with the 3D flowfield of the baseline configuration to be used as reference. The network then predicts the 3D flowfield for the configurations of interest, consisting of the contours for all variables at the axial locations of interest. The outputs are averaged to obtain the relevant radial profiles and integral quantities, which are then used to calculate stage-wise performance first, and then the overall performance using relevant thermodynamic equations. 

\subsubsection{Model architecture} \label{sec:network}

The architecture shown in \autoref{fig:architecture}, is a variant of the one originally proposed by the authors in previous work \cite{Bruni2023} \cite{Bruni_Maleki_Krishnababu_2025}. The input variables consists of three separate tensors: the manufacturing and build variables in a tensor of size $(n_{build} \times n_{rows})$, blade geometry design parameters in a tensor of size $(n_{geom}\times n_{rows} \times n_{sect})$, operating conditions in a sensor of size $n_{op}$. A pre-processing step converts the input tensors to a shape matching the 3D flow field, which is then concatenated to the flow field of the baseline case in a tensor of size $(n_{vars} \times n_{rows} \times n_{rad} \times n_{tang})$. The baseline flow field is provided as an input in the first iteration of the inner loop, while for later iterations the output of the predictions are fed back at the input. The model is based on a 3D U-Net architecture with double residual convolutional blocks. Each convolutional block consists of a 3D convolution layer, followed by batch normalization and an \textit{ELU} activation function. A residual connection is implemented between the first convolution and the second activation function. The use of residual connections \cite{ResUnet} was found to significantly improve the training for deeper networks, as it facilitates the propagation of the information across the network. The number of axial locations is fixed to $n_{rows}$, and each down-sampling convolutions and up-sampling transposed convolutions uses a stride of $(1,2,2)$, which retains the dimension consisting of the inter-row locations, which is physically more meaningful than traditional constant strides across dimensions. The network can learn the complex relationships between the different inter-row locations, as the flow field at a given location is defined by the upstream and downstream conditions, as well as the input features representative of that blade-row (i.e. geometry parameters, tip clearance and surface roughness). Both up and down samplings are followed by a double convolutional block at each layer, while the up-sampling are also followed by attention blocks, which focus on the most relevant parts of the input tensor, enhancing the feature representation. By selectively weighting different regions of the input, the network can prioritize important features while ignoring less relevant ones, thus generalising better to unseen data by dynamically adjusting the focus based on the input, as well as capturing long-range dependencies within the input data. The output of the network is a tensor of size $(n_{vars} \times n_{rows} \times n_{rad} \times n_{tang})$, which encompasses the entire flow field. The encoding section followed by the decoding section, along with the skip connections, enables the network to predict both low-level and high-level features in the flow field. More complex models such as U-Net++ \cite{Zhou2018} were considered but did not lead to a significant improvement in the accuracy of the predictions, and were not considered worth the increased computational cost for the applications of interest.

\subsubsection{Iterative networks}
In the C(NN)FD framework, the baseline contours tensor $y_{baseline}$ is concatenated along the channel axis ($\oplus$) with input variables tensor $x$, which are then passed through the network $f$ with weights $\Theta$ to provide the output $y_0$ as in \autoref{eq:baseline}. This approach has been demonstrated to improve prediction accuracy \cite{Fesquet2024}, as leveraging the baseline contours as a reference provides valuable contextual information to the model, reformulating the problem in terms of the deviations in input features relative to the baseline. The network can then better understand the relationships between the input variables and the resulting flow field, leading to improved feature representation, reduced prediction errors, and enhanced generalization. 
\begin{equation}
   y_0 = f_{\Theta}(x \oplus y_{baseline})
   \label{eq:baseline}
\end{equation}
While providing a baseline tensor as an input to the network is known to improve the accuracy of the predictions, the resulting change in architecture also provides the option to fed through the output of the network at the input, in an iterative network \cite{Mosinska2018}. For each epoch, the output of the network is fed through the input for a defined number of inner loops $N$ as in \autoref{eq:iterative}, defining a supervised learning approach in which each intermediate output is used to sum the overall loss, as in \autoref{eq:iterative_loss} using the node-wise loss function $\mathcal{L}$ against the ground truth $Y$. 
\begin{equation}
   y_i = f_{\Theta}(x \oplus y_{i-1})
   \label{eq:iterative}
\end{equation}
\begin{equation}
   \mathcal{L}_N = \sum_{i=1}^N \mathcal{L}_i(y_i, Y)
   \label{eq:iterative_loss}
\end{equation}
The network can then further refine its predictions and improve their accuracy compared to a single forward pass. While the increases number of $xN$ forward passes increases computational cost, this is partially offset by a faster convergence rate and a reduced number of epochs.

\subsubsection{Uncertainty Estimation}

In addition to the improved prediction accuracy that the iterative model offers, the internal loops provide the opportunity to quantify the associated uncertainty across the iterations, as the convergence rate of the outputs is correlated with the uncertainty of the predictions \cite{Durasov2024}. This is achieved by performing multiple forward passes through the network and calculating the standard deviation of the predictions, as in \autoref{eq:uncertainty} \cite{Durasov2024}. Both the epistemic and aleatoric uncertainties associated with the predictions are estimated, providing a quantifiable measure of the confidence in resolving the relevant flow features. For the applications of interest in this work, epistemic uncertainty is considered to be the most significant, due to the limited size of the datasets available, and the high likelihood of out-of-distribution data. In the case of operating conditions in which the CFD model is know to have higher uncertainty, such as near stall conditions, aleatory uncertainty is also expected to be significant. High uncertainty predictions can then be addressed by gathering more data representative either the in-distribution with aleatoric noise candidates, such as near stall conditions, or of the out-of-distribution candidates, such as combinations of the input features for which enough training data is not provided. 

\begin{equation}
   U^i = Var(\{y_1^i, y_2^i, \dots , y_N^i\})
   \label{eq:uncertainty}
\end{equation}

While several techniques can be employed to estimate the uncertainty in the predictions of a machine learning model, Deep Ensembles \cite{Laks2017} are currently the most popular approach due to their reliable uncertainty estimations. However, training multiple networks and performing several forward passes significantly increases computational cost. Iterative networks have been demonstrated to provide comparable accuracy, with a significantly lower computational cost, and no change in architecture required \cite{Durasov2024}. Estimating uncertainty is particularly useful for identifying regions of the flow field where the model is less certain, and regions of the design space that might require additional data. Quantifying prediction confidence is crucial for decision-making in industrial applications, providing a metric to assess the reliability of the model's predictions and identify potential risks. 

\subsubsection{Multi-dimensional physical loss function}
The framework uses a physics-informed approach, in which the relevant physical equations are embedded as soft constrains in the loss function \cite{Raissi2019}, which is defined to ensure that the predictions are accurate and physically meaningful. The loss is calculated using a combination of different components, including the primary flow field, its spatial gradient \cite{Fesquet2024}, the radial distributions, and integral quantities as in \autoref{eq:physical_loss}. 

\begin{equation}
   \mathcal{L} = \mathcal{L}_{contours} + \mathcal{L}_{grad} + \mathcal{L}_{radial} + \mathcal{L}_{integral}
   \label{eq:physical_loss}
\end{equation}

For each loss component, the difference between the predicted and ground truth flow field variables is used to calculate the \textit{Huber} loss function, which computes the mean squared error (MSE) below a defined threshold, or the mean absolute error (MAE) above it to achieve better robustness to outliers. The primary component of the loss is calculated based on the predicted flow field variables and their 3D spatial gradients at each axial location. The model also calculates the radial distributions and integral quantities at each plane through area averaging during training, and these are used to define the corresponding loss functions. These additional components of the loss ensure that the model's predictions of the flow field are consistent with the governing physical principles, and significantly improve the overall performance predictions, which are highly sensitive to small errors in the inter-row planes. 

\subsubsection{Training}

The network is trained using the \textit{AdamW} optimizer, and the datasets are divided into three sets for each task: 70\% training data, 20\% validation data, and 10\% test data. The validation set was utilized to evaluate model performance and perform hyper-parameter tuning, while the test set was reserved only for a final assessment. This ensures that the results are generalisable to unseen cases, and will lead to comparable performance also when deployed in a production environment. The dataset is split in a stratified manner, dividing the data in 10 bins using a Principal Component Analysis (PCA) of the input features. This ensures that each dataset has a comparable distribution. The datasets are scaled based on the minimum and maximum value of each variable in the training set. The training converges in around 500 epochs depending on the dataset, with a batch size of 20 and an initial learning rate of 0.01. A learning rate scheduler was implemented, which halved the learning rate with a patience of 20 epochs. To avoid overfitting, early stopping with a patience of 50 epochs was implemented. The training time for a new model from scratch is around 1 hour on a single Tesla T4 GPU, for a dataset containing $\sim 400$ CFD cases. This is reduced to less than 20 minutes when re-training the network when data becomes available, thus making the architecture scalable for industrial applications. The execution time for inference is less than 1 second, making the predictions of the deployed model effectively real-time for the application of interest.

\section{Machine learning applications to compressor aerodynamics}
The C(NN)FD framework can support the workflow of CFD engineers working on compressor aerodynamics, spanning across various stages of the design and analysis processes. These can include accelerating design iterations, by providing rapid performance and flow field predictions, or as part of design optimisations as a low-fidelity surrogate models. The framework can also quantify the sensitivity of performance metrics to variations in design parameters such as blade geometry, and manufacturing and build variations such as tip clearance and surface roughness. By identifying trends and trade-offs, appropriate tolerance limits can be set to ensure consistent performance across different builds. The framework can also be used during the manufacturing and build process to assess the performance of specific engine builds, and identify ones that would not meet contractual performance requirements. The following sections present three example applications to typical design and analysis activities: the effect of design variations, the off-design assessment of manufacturing and build variations, and a demonstration of transfer learning to predict the effect of manufacturing and build variations on a different compressor design. All values presented are non-dimensionalized with respect to the results obtained from the corresponding baseline.

\subsection{Effect of design variations}

\begin{figure}[!htbp]
   \includegraphics[width=0.42\textwidth]{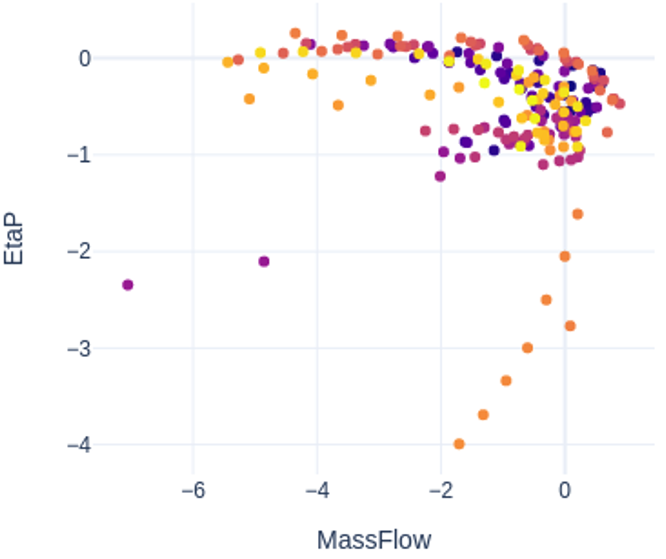}
   \caption{CFD Database resulting from the design activity}
   \label{fig:design_database}
\end{figure}

The C(NN)FD framework is demonstrated on a typical design activity, in which a datum compressor geometry was re-designed targeting an efficiency improvement at a given operating condition. This was performed through an automated design process in which all the rotors and stators airfoil designs were allowed to vary, without any specific constraints other than mechanical design considerations. For each blade-row, the geometry is defined at 5 radial sections, and parametrised for a given airfoil family using the following design variables: inlet angle, camber angle, thickness, chord and pitch. The blade metal angles and chord were considered as design variables, while thickness and pitch were fixed. An overview of the dataset with the polytropic efficiency and mass-flow of all the design iterations is presented in \autoref{fig:design_database}, consisting of 200 geometries and CFD solutions, suggesting that candidates with improved efficiency compared to the baseline resulted in lower mass-flow. The dataset also includes candidates generated in the initial stages of the design process which significantly underperform compared to the baseline.

\begin{figure}[!htbp]
       \includegraphics[width=0.45\textwidth]{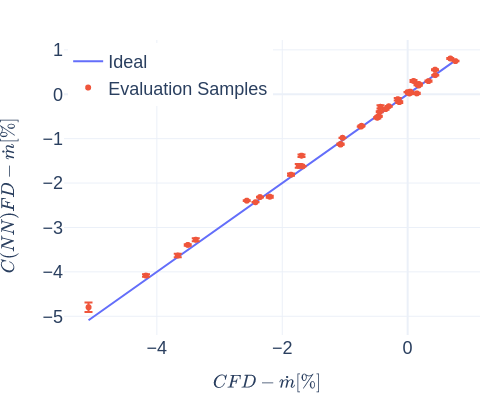}
       \caption{Effect of design variations: $\dot{m}$ comparison}
       \label{fig:Overall_Mass}
\end{figure}

\begin{figure}[!htbp]
       \includegraphics[width=0.45\textwidth]{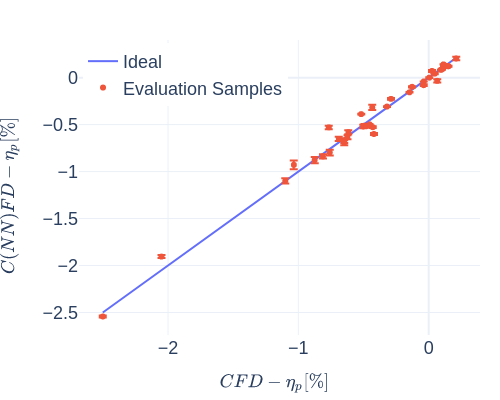}
       \caption{Effect of design variations: $\eta_p$ comparison}
       \label{fig:Overall_EtaP}
\end{figure}

The aim is to demonstrate that a machine learning model can be trained on the results of a preliminary design activity, and can be used as a low-fidelity model to drive further design iterations, or an optimisation. The overall performance predictions of the trained model are summarised in \autoref{table:OverallPerformance}, and an overview of the comparison is provided in  \autoref{fig:Overall_Mass} and \autoref{fig:Overall_EtaP}. 

\begin{table}[!htbp]
   \centering    
   \caption{Effect of design variations: performance metrics}
   \begin{tabular}{c|ccc}
    Var       & $R^2$ & $MAE \%$   & $\Delta dataset \%$   \\ \hline
    $\dot{m}$ & 0.9947 & 0.0758         &  10.81             \\      
    $\eta_p$  & 0.9847 & 0.0461          &  3.01              \\ 
   \end{tabular}
\label{table:OverallPerformance}
\end{table}


\begin{figure}[!h]
   \includegraphics[width=0.49\textwidth]{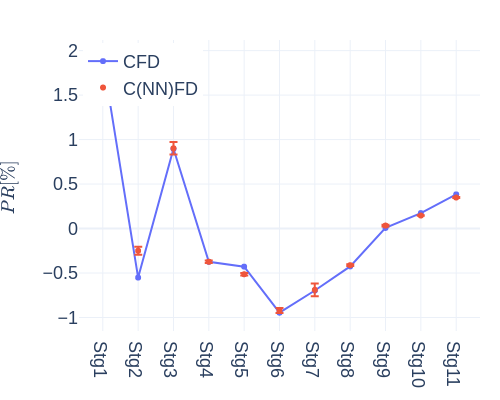}
   \caption{Effect of design variations: PR Stage-wise distribution}
   \label{fig:Stagewise_PR}
\end{figure}

\begin{figure}[!h]
   \includegraphics[width=0.49\textwidth]{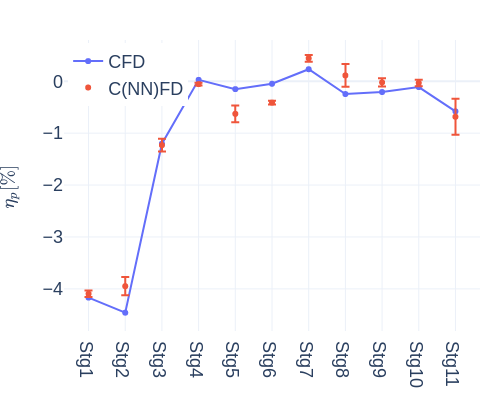}
   \caption{Effect of design variations: $\eta_p$ Stage-wise distribution}
   \label{fig:Stagewise_EtaP}
\end{figure}

Excellent agreement is observed for both the mass-flow and efficiency predictions, with a coefficient of determination $R^2$ close to 1 and a Mean Absolute Error (MAE) two orders of magnitude smaller than the range of the dataset $\Delta dataset \%$ for each variable. The range of the overall performance variables is provided for reference, to relate the errors in the machine learning predictions to the variations considered in the dataset for the application of interest. Even design candidates with significantly lower efficiency and mass-flow than the baseline, representing out-of-distribution designs, are predicted accurately. The predictions could be further improved by providing more training data as the design activity progresses, to better characterise the design space. The stage-wise distributions for the design candidate with the greatest discrepancy in overall performance between the predictions and ground truth, an out-of-distribution candidate with an efficiency drop of more than $2\%$ compared to the baseline, are shown in \autoref{fig:Stagewise_PR} for pressure ratio and \autoref{fig:Stagewise_EtaP} for polytropic efficiency. C(NN)FD accurately predicts the trend for each stage relative to the baseline with excellent agreement. Only minor discrepancies are noticeable, with the variations being either slightly over-predicted or under-predicted depending on the stage, due to the low availability of representative data. 


\begin{figure}[!htbp]
       \includegraphics[width=0.41\textwidth]{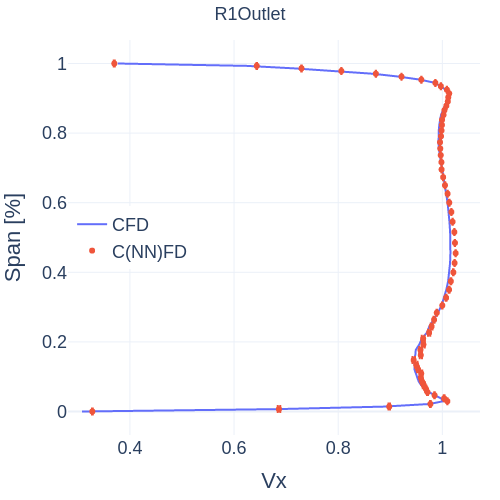}
      \caption{Effect of design variations: Rotor-1 $V_x$ radial profile}
      \label{fig:Worst_Radial_1}
\end{figure}

\begin{figure}[!htbp]
   \includegraphics[width=0.41\textwidth]{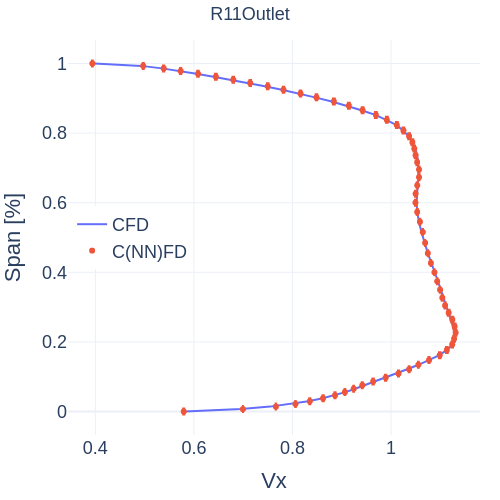}
  \caption{Effect of design variations: Rotor-11 $V_x$ radial profile}
  \label{fig:Worst_Radial_11}
\end{figure}

 Even though the front-stages present a significantly lower efficiency than the baseline, the model is still able to capture the trend accurately. This is achieved through the multi-dimensional physical loss function which includes the integral quantities at each inter-row location in the supervised training. While the row-wise distributions of the flow variables are not presented for conciseness, they present similar agreement to that of the stage-wise distributions. The results for the radial profiles of axial velocity for Rotor-1 and Rotor-11 are presented in \autoref{fig:Worst_Radial_1} and \autoref{fig:Worst_Radial_11}, while all the other stages and variables exhibit similar level of agreement and are not presented for conciseness. The agreement in the radial distribution predictions is generally good, except for some localised discrepancies such as Rotor-1 exhibiting some un-physical discontinuities towards 10\% span.

Future work to address this issue could include considering the gradients of the radial profiles as additional loss function, or  providing more training data. The prediction error for Rotor-1 can be physically explained by the significant effect that design variations can have on the flow field of transonic stages, which are characterised complex shock patterns and sharp gradients in the flow field. The predictions for Rotor-11 on the other hand shows much better agreement, as the effect of design variations on the flow field is more linear and the database available included fewer out-of-distribution instances.

\begin{figure}[!h]
      \includegraphics[width=0.49\textwidth]{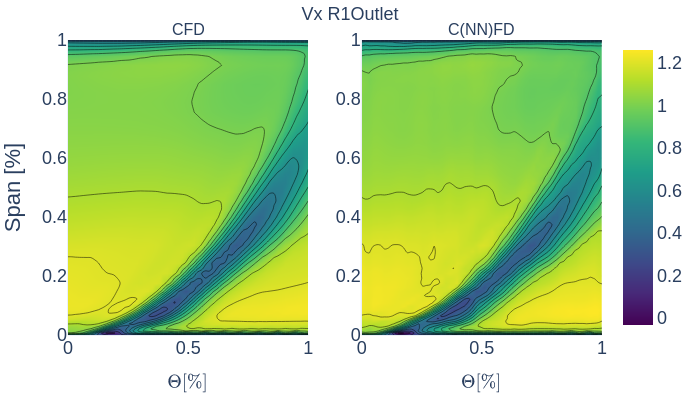}
      \caption{Effect of design variations: $V_x$ contour comparison, Rotor 1 outlet}
      \label{fig:Worst_rotor_1}
   \end{figure}

\begin{figure}[!h]
     \includegraphics[width=0.49\textwidth]{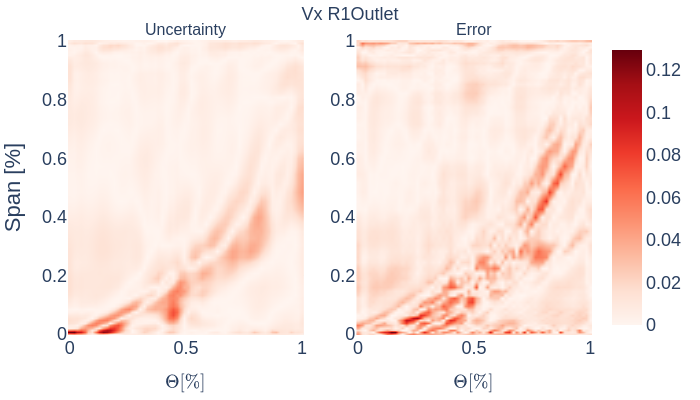}
     \caption{Effect of design variations: $V_x$ error and uncertainty, Rotor 1 outlet}
     \label{fig:Worst_rotor_1_error}
\end{figure}

\begin{figure}[!h]
   \includegraphics[width=0.49\textwidth]{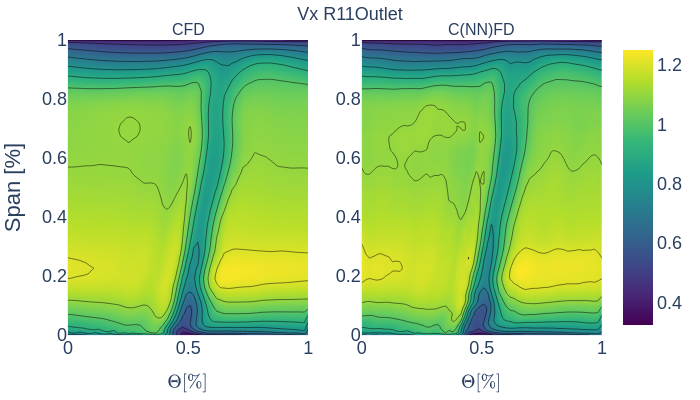}
   \caption{Effect of design variations: $V_x$ contour comparison, Rotor 11 outlet}
   \label{fig:Worst_rotor_11}
\end{figure}

\begin{figure}[!h]
  \includegraphics[width=0.49\textwidth]{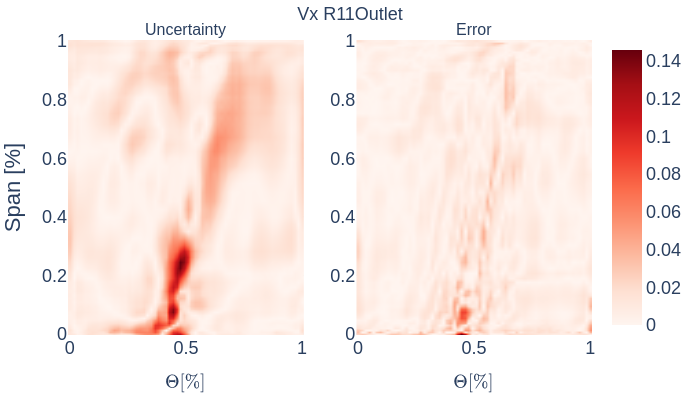}
  \caption{Effect of design variations: $V_x$ error and uncertainty, Rotor 11 outlet}
  \label{fig:Worst_rotor_11_error}
\end{figure}

The axial contours are presented for Axial Velocity $V_x$ in \autoref{fig:Worst_rotor_1} at Rotor-1 outlet and \autoref{fig:Worst_rotor_11} at Rotor-11 outlet, with CFD predictions in the left and C(NN)FD predictions on the right. The corresponding uncertainty predictions and relative errors are provided in \autoref{fig:Worst_rotor_1_error} and \autoref{fig:Worst_rotor_11_error}. All the other stages and variables exhibit similar level of agreement and are not presented for conciseness.


\newpage 
The $V_x$ comparison presented is used to judge the quality of the machine learning model in predicting the relevant aerodynamic flow features. This includes for example the regions of low axial momentum associated with the airfoil wakes, the tip clearance flows or separated flows. The predictions generated by \textit{C(NN)FD} exhibit excellent agreement with the ground truth, with the primary flow features being accurately reproduced. For both front and rear stages, the main regions of uncertainty and predictions errors, are the wake region, while the end-wall flow features tend to be captured accurately. The errors are localised to regions with steep gradients, and are identified by the corresponding uncertainty estimates. This is due to the wide range of design variables that was included in the design activity, which led to significantly different airfoil geometries, especially for the transonic stages. The C(NN)FD contours are generally less smooth than the CFD ones, and some artificial discontinuities in the flow field are present even if the spatial gradients are already included in the multi-dimensional physical loss function. Future work will research additional physical constraints that can be imposed during training to achieve more physical results.

The uncertainty estimates have been demonstrated to qualitatively identify the flow features in the contours which are not resolved accurately due to lack of relevant examples in the training dataset. However, the comparison between predictions and ground truth for the integral quantities suggest that for some specific instances the uncertainty estimates could be under-predicted compared to the errors. Future work will focus on benchmarking different uncertainty estimation methods, based on their relative accuracy and trade-off with computational cost.

\subsection{Off-design assessment of manufacturing and build variations}

The C(NN)FD framework is then demonstrated on the modelling of the effect of manufacturing and build variations on the off-design performance of a given compressor design. The CFD data used to train the model was discussed in previous publications with regards to the effect of manufacturing and build variations on the compressor aerodynamic performance \cite{Bruni2024CFD} and on the machine learning modelling of the effect of manufacturing and build variations on the design point performance \cite{Bruni_Maleki_Krishnababu_2025}. 

\begin{figure}[!h]
   \includegraphics[width=0.49\textwidth]{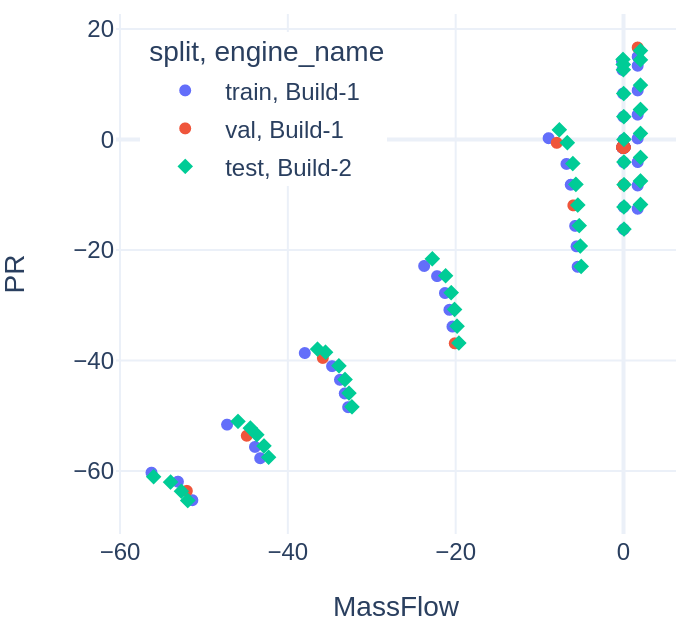}
   \caption{Train, validation and test sets for the assessment of manufacturing and build variations at off-design conditions}
   \label{fig:compressor_map_manufacturing_build}
\end{figure}

\begin{figure}[!h]
   \includegraphics[width=0.49\textwidth]{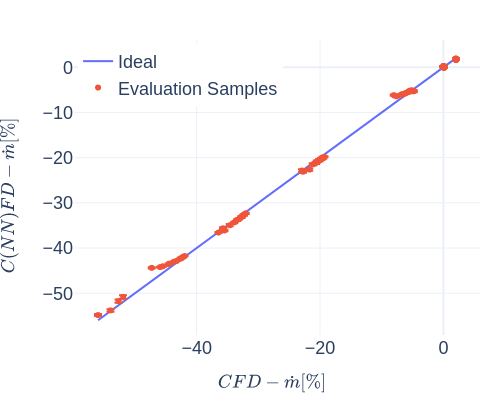}
   \caption{Off-design assessment: $\dot{m}$ comparison}
   \label{fig:off_des_mass}
\end{figure}

\begin{figure}[!h]
   \includegraphics[width=0.49\textwidth]{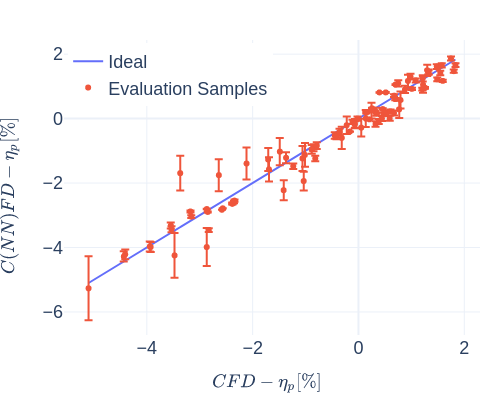}
   \caption{Off-design assessment: $\eta_p$ comparison}
   \label{fig:off_des_eff}
\end{figure}

\newpage
An overview of the dataset is presented in \autoref{fig:compressor_map_manufacturing_build} with the corresponding split in train, validation and test sets, consisting of a total of 500 CFD solutions available. The dataset is composed of two subsets. First, 400 CFD calculations include different manufacturing and build variations of a given geometry at design point \autoref{fig:compressor_map_manufacturing_build}. The tip clearance and surface roughness were varied for each row within out-of-tolerance conditions up to 50\% larger and tighter than the drawing specifications. This range is significantly larger than the tolerance typically specified for gas turbine applications, and was selected to demonstrate the robustness of the methodology to out-of-tolerance cases. The input variable space was sampled using latin-hypercubic sampling. Secondly, the corresponding compressor map including 50 additional CFD solutions for the same compressor design consisting of 7 different speedlines between 80\% and 105\% speed for a specific compressor build, namely Build-1, was included. The tip clearance and surface roughness values for this specific engine build are within drawing tolerances. The dataset is then split between train and validation set, and is used to train C(NN)FD. The trained model is then used to predict the performance of the test set, which consists of a different compressor build, namely Build-2, across the whole operating range at 50 different operating points. In this example, the two engine builds have the same airfoil geometries, but different tip clearances and surface roughness, with Build-1 having higher surface roughness when compared to Build-2. It should be noted that this presents a significant challenge aerodynamically, due to the wide operating range considered across the whole compressor map, up to the numerical stability of the CFD model. The overall performance predictions from C(NN)FD are compared with the ground truth for the test set with regards to mass-flow in \autoref{fig:off_des_mass} and efficiency in \autoref{fig:off_des_eff}, and summarised in \autoref{table:OverallPerformance_manufacturing_build}. 


\begin{table}[!htbp]
   \centering    
   \caption{Off design assessment: performance metrics}
   \begin{tabular}{c|ccc}
    Var       & $R^2$ & $MAE \%$   & $\Delta dataset \%$   \\ \hline
    $\dot{m}$ & 0.9989    &     0.4087           &  58.01             \\      
    $\eta_p$  & 0.9536    &      0.2685          &  6.94              \\ 
   \end{tabular}
\label{table:OverallPerformance_manufacturing_build}
\end{table}

\newpage
While the $MAE$ is higher than for the previous application, due to the datasets being more heterogeneous, as demonstrated by the wider dataset range for both variables, the $R^2$ is still close to 1s. The mass-flow predictions are excellent, and for each cluster corresponding to every speedline, the discrepancy increases only towards near stall conditions, which are notoriously challenging to resolve and have higher aleatoric uncertainty. Likewise, the efficiency predictions also present good agreement, with greater discrepancies observed for the near stall conditions, which are also flagged by the model as with higher uncertainty. 

\begin{figure}[!h]
   \includegraphics[width=0.49\textwidth]{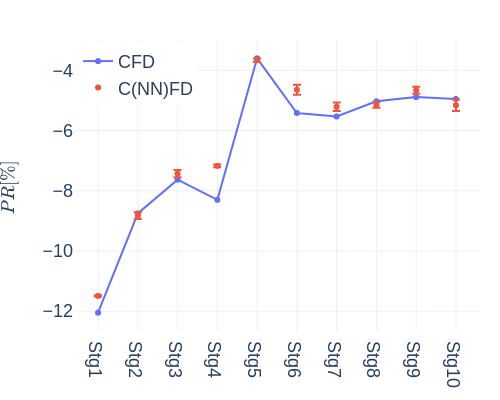}
   \caption{Off design assessment: PR Stage-wise distribution}
   \label{fig:off_des_stg_eff}
\end{figure}

\begin{figure}[!h]
   \includegraphics[width=0.49\textwidth]{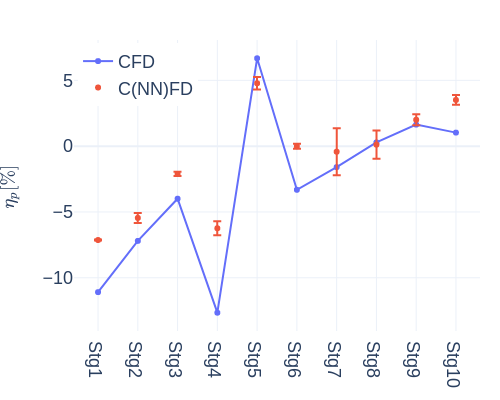}
   \caption{Off design assessment: $\eta_p$ Stage-wise distribution}
   \label{fig:off_des_stg_pr}
\end{figure}

The corresponding stage-wise distributions are presented in \autoref{fig:off_des_stg_eff} and \autoref{fig:off_des_stg_pr} for the case with the greatest error, representative of near-stall operating conditions. Even for challenging conditions aerodynamically, they predictions were found to match closely, with the relative trends to the baseline accurately captured. This demonstrates the potential to apply the methodology for off-design assessments, where limited data is available. Due to the complexity and non-linearity of the flow field, the predictions are expected to improve as more training data is provided representative out-of-distribution cases.

\subsection{Transfer learning to a different compressor design}
\begin{figure}[!htbp]
   \includegraphics[width=0.45\textwidth]{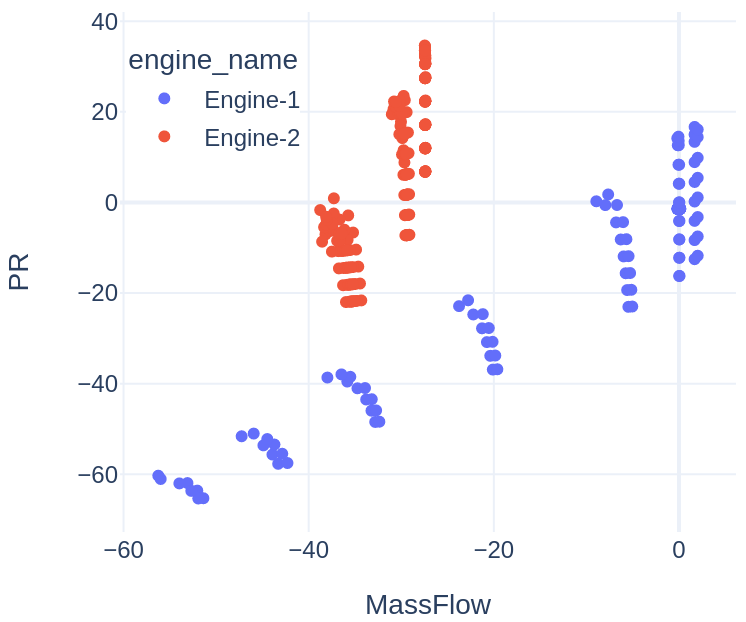}
   \caption{CFD Database used for the transfer learning assessment from Engine-1 to Engine-2}
   \label{fig:compressor_map_transfer}
\end{figure}

The transfer learning approach is demonstrated by fine-tuning the model presented in the previous section for a given compressor design (Engine-1), to a different dataset consisting of a different compressor design (Engine-2) as shown in  \autoref{fig:compressor_map_transfer}.

\begin{figure}[!htbp]
   \includegraphics[width=0.49\textwidth]{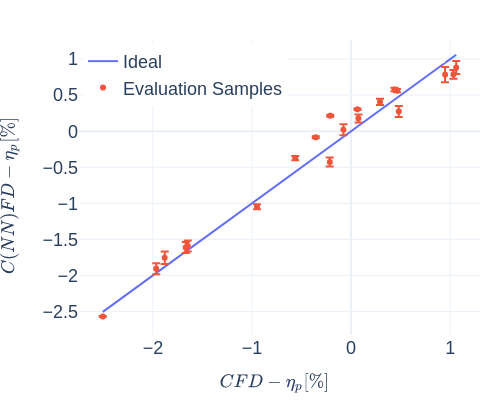}
   \caption{Transfer learning application: $\eta_p$ comparison}
   \label{fig:transfer_eff}
\end{figure}

\begin{figure}[!htbp]
   \includegraphics[width=0.49\textwidth]{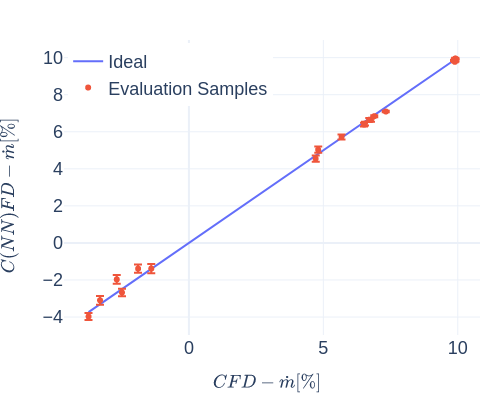}
   \caption{Transfer learning application: $\dot{m}$ comparison}
   \label{fig:transfer_mass}
\end{figure}

The two compressors designs have significantly different gas paths, airfoil geometries, design rotational speed, and VGV schedules. Three speedlines, are considered for the assessment: 95\%, 100\% and 105 \% speed. This test-case was selected to demonstrate the potential of transfer learning, as training a model from scratch using only the dataset available for Engine-2 would not provide sufficient accuracy. The aim is to predict the performance of different compressor builds across the whole operating range, using the pre-trained weights and the CFD data for a given compressor build for training. The dataset consists of 200 CFD calculations. The compressor map is available for 8 different compressor builds, each with different surface roughness within drawing specifications. An overview of the result is presented in \autoref{fig:transfer_eff} and \autoref{fig:transfer_mass} and the summary in \autoref{table:OverallPerformance_transfer}. 
 

\begin{table}[!htbp]
   \centering    
   \caption{Transfer learning assessment: performance metrics}
   \begin{tabular}{c|ccc}
    Var       & $R^2$ & $MAE \%$   & $\Delta dataset \%$   \\ \hline
    $\dot{m}$ & 0.9977    &     0.1634          &  14.97             \\      
    $\eta_p$  & 0.9699    &     0.1575          &  3.62              \\ 
   \end{tabular}
\label{table:OverallPerformance_transfer}
\end{table}
 
Without the use of transfer learning, the machine learning model would not be able to learn accurately the effect of different surface roughness values on the flow field across the whole compressor maps, as only 8 examples for each operating points are available. The use of transfer learning, allows to transfer the patterns of the effects of different surface roughness across the compressor map for a given engine, and to use that as a base for the training on a more limited dataset for a different engine.  demonstrate that through the use of transfer learning, excellent agreement between the C(NN)FD predictions and the CFD results can be achieved even with a very limited amount of training data, across a wide range of operating conditions. This unlocks the potential to re-use pre-trained machine learning models developed for a given test case in which sufficient data was available, and to re-train them for a different application for which only a limited dataset is available.

\section{Conclusion}
This paper presents a novel deep learning framework for predicting the flow field and aerodynamic performance of multi-stage axial compressors, which is also potentially applicable to any type of turbomachinery. A physics-based dimensionality reduction unlocks the potential for flow-field predictions for large-scale domains, re-formulating the regression problem from an unstructured to a structured one and reducing the number of degrees of freedom. The relevant physical equations are used to define a multi-dimensional physical loss function, which considers the flow field, gradient of the flow field, radial distributions and integral quantities. Compared to “black-box” approaches, the proposed framework has the advantage of physically explainable predictions of overall performance, as the corresponding aerodynamic drivers can be identified on a 0D/1D/2D/3D level. An iterative architecture is employed, improving the accuracy of the predictions, as well as estimating the associated uncertainty at each level of the predictions. A quantifiable measure of the confidence in resolving the relevant flow features is provided, without a significant increase in computational cost. The model's generalizability is demonstrated across a series of dataset including manufacturing and build variations, varying geometries, compressor designs and operating conditions, representative of typical aerodynamic assessments, making it a valuable tool for industrial applications in turbomachinery.


\section*{Acknowledgments}
The authors would like to thank Siemens Energy Industrial Turbomachinery Ltd. for allowing the publication of this research, as well as Richard Bluck and Roger Wells of Siemens Energy for their support and comments. This work was supported and funded by Siemens Energy Industrial Turbomachinery Ltd.

\section*{Permission for Use}
The content of this paper is copyrighted by Siemens Energy Global GmbH \& Co. KG.  Any inquiries regarding permission to use the content of this paper, in whole or in part, for any purpose must be addressed to Siemens Energy Industrial Turbomachinery Limited, directly. 

\bibliographystyle{asmeconf}  
{\footnotesize
\bibliography{GT2025-151295}}

\end{document}